# Improved Heat Dissipation in $CsPbBr_3$-hBN Heterostructures

Liudmila Starodubtceva, [a] Ahmed Kadid, [b] Naho Kurahashi, [b] Cedric Kreusel, [b] Jasper Ruhkopf, [ad] Sebastian Lukas, [a] Ulrich Plachetka, [d] Ralf Heiderhoff, [bc] Thomas Riedl, [bc] Maryam Mohammadi, *[d] Max C. Lemme *[ad]

[a] Chair of Electronic Devices, RWTH Aachen University, Otto-Blumenthal-Str. 25, 52074 Aachen, Germany.

[b] Institute of Electronic Devices, University of Wuppertal, Rainer-Gruenter-Str. 21, 42119 Wuppertal, Germany.

[c] Wuppertal Center for Smart Materials & Systems (CM@S), University of Wuppertal, Rainer-Gruenter-Str. 21, 42119 Wuppertal, Germany

[d] AMO GmbH, Otto-Blumenthal-Str. 25, 52074 Aachen, Germany.

*Corresponding Authors: max.lemme@eld.rwth-aachen.de, mohammadi@amo.de

**Abstract**

Metal halide perovskite semiconductors are promising materials for optoelectronic and photonic devices, including solar cells and next-generation coherent light sources. However, their low thermal conductivity limits the practical operation of devices under high excitation levels. Integrating thermally conductive, large band-gap two-dimensional (2D) materials into perovskite devices could suppress heat accumulation, while preserving their optical properties. Here, we show that planar hot-pressed (PHP) cesium lead bromide ($CsPbBr_3$) thin films capped with few-layer 2D hexagonal boron nitride (hBN) are less affected by laser-induced heating under high-power continuous-wave excitation than uncapped perovskite samples. A large-scale semidry transfer method was developed to integrate 2D hBN onto PHP $CsPbBr_3$ thin films. The process is chemically and thermally compatible with perovskites. Microscopic and spectroscopic analyses show that the hBN capping layer does not alter the morphology and optical properties of the perovskite thin film. The PHP $CsPbBr_3$-hBN heterostructure exhibits a thermal conductivity of 3 W/(m·K), approximately seven times higher than that of the bare perovskite films of 0.45 W/(m·K). Heat diffusion simulations confirm enhanced heat dissipation in the heterostructure relative to bare perovskite films. Our experiments demonstrate an effective approach to enhancing heat dissipation in perovskite devices using transparent, thermally conductive 2D materials.

**Introduction**

Metal halide perovskite semiconductors have demonstrated high photoluminescence quantum yields and optical gain, making them strong candidates for next-generation coherent light sources. [1,2] However, their low thermal conductivity poses a challenge for LEDs, solar cells, and lasing devices. Improving heat dissipation is essential for reducing efficiency roll-off and enhancing operational stability.[3–6]

Several strategies have been explored to mitigate thermal limitations in perovskite-based devices[7]. Low-thermal-conductivity glass substrates have been replaced by sapphire,[8] and thermally conductive two-dimensional (2D) materials such as surface-terminated titanium carbide ($Ti_3C_2T_x$) have been integrated into perovskite layers to reduce the temperature of device during operation. [9] In addition, introducing ultrathin hexagonal boron nitride (hBN) at perovskite grain boundaries has been shown to reduce the operating temperature when combined with a copper (Cu) heat spreader.[10] hBN is characterized by strongly anisotropic thermal transport, with thermal conductivity values of up to 8 W/(m·K) in the cross-plane direction and up to 484 W/(m·K) in the in-plane direction.[11–13] Furthermore, its wide band gap of approximately 5.8-6.4 eV renders hBN optically transparent across the emission range of most halide perovskites.[14] These properties make hBN a promising material for thermal management in halide perovskite optoelectronic devices.

Furthermore, several studies have investigated the use of hBN as an encapsulation layer. [15,16] In these works, hBN primarily acts as a chemically inert, optically transparent, and mechanically robust diffusion barrier that protects the perovskite from environmental degradation.[15,17] Despite these advantages, the integration of hBN onto perovskite thin films remains challenging because of the pronounced sensitivity of perovskites to common processing conditions. Conventional transfer procedures for 2D materials frequently rely on long exposure to temperatures up to 180 °C,[18] polar solvents,[19] and etchants,[20] which are incompatible with perovskite materials. Dry transfer techniques, such as Polydimethylsiloxane (PDMS)-assisted exfoliation, avoid liquid-phase processing but often require additional support layers and precise handling, limiting scalability.[21,22]

In this work, we demonstrate that encapsulating planar hot-pressed (PHP) cesium lead bromide ($CsPbBr_3$) thin films with few- layer 2D hBN via a semidry transfer significantly improves heat dissipation relative to uncapped perovskite films. The improved thermal management is evidenced by a reduced laser-induced affected area under high-power continuous wave excitation and confirmed by optical spectroscopy. The structural and optical quality of the perovskite films was assessed using scanning electron microscopy (SEM), amplified spontaneous emission (ASE), and X-ray diffraction (XRD). The thermal properties of the resulting heterostructures were investigated via scanning near-field thermal microscopy (SThM).[23] Heat transport simulations employing literature-reported thermal parameters for hBN, silicon dioxide ($SiO_2$), and silicon (Si) were used to analyze the cooling dynamics of the heterostructure.

**Results and discussion**

The fabrication of the perovskite-2D material heterostructures involved solution-based deposition of $CsPbBr_3$ thin films followed by a semidry transfer of hBN optimized for perovskite compatibility. The $CsPbBr_3$ thin films used in this study were spin-coated from a solution of cesium bromide (CsBr) and lead bromide ($PbBr_2$) in dimethyl sulfoxide (DMSO) onto silicon-silicon oxide ($Si$-$SiO_2$) and glass substrates, followed by post-annealing at 100 °C and PHP (see Experimental section for details).[2] Briefly, PHP is a variant of thermal imprint using a planar stamp. It has previously been shown to be a powerful technique for post-treating various halide perovskite thin films, affording significantly reduced surface roughness and improved optical properties.[2,24] The smooth surface of the perovskite enables reliable transfer of 2D materials onto it.

The hBN transfer process, illustrated in Fig. 1, followed a reported method[18] and avoided processing conditions incompatible with perovskite materials. First, chemical vapor deposition (CVD)-grown hBN on Cu foil was coated with poly(methyl methacrylate) (PMMA) to support and protect the 2D layer during transfer. Subsequently, a plastic frame was attached to avoid any mechanical contact with the hBN-PMMA surface. To release the hBN, the Cu substrate was etched by a hydrogen peroxide and hydrochloric acid solution. Subsequently, the hBN-PMMA frame was transferred onto the perovskite film under inert conditions and briefly heated to temperatures of up to 140 °C, at which the PMMA softens, allowing hBN layer to conform to the perovskite surface. After transfer, the plastic frame was removed, and the PMMA support layer was dissolved in chlorobenzene, a nonpolar solvent benign to the perovskite films.

Figure 2a shows an optical image of a PHP $CsPbBr_3$ sample prepared with partial hBN coverage, such that the hBN-covered and uncovered regions underwent identical processing steps for direct comparison. SEM images (Fig. 2b, c) show no change in the perovskite morphology prior to and after hBN layer transfer. The surface image acquired under the same contrast, brightness, and accelerating voltage settings appears brighter in Fig. 2c. This effect is attributed to higher surface charging, as hBN is a dielectric material. ASE measurements reveal no change in the threshold, indicating that the wide-bandgap hBN layer is sufficiently thin, optically transparent, and thus does not affect the ASE performance of PHP $CsPbBr_3$ (Fig. 2d).

The XRD patterns of PHP $CsPbBr_3$ and PHP $CsPbBr_3$-hBN show characteristic peaks of the pure orthorhombic $CsPbBr_3$ phase (reference code: 00-018-0364) (Fig. 2e). The hBN (002) peak at $2\theta \approx 26.7°$ overlaps with an intense $CsPbBr_3$ diffraction peak, preventing its unambiguous identification. These results indicate that the transfer process does not affect the crystallinity of the underlying perovskite.

SThM, an AFM-based scanning probe technique, was employed to investigate the thermal transport properties of the samples (Fig. 3a). Owing to the low thermal conductivity of the underlying perovskite layer, heat dissipation in the heterostructure occurs predominantly laterally. The topography images (Fig. 3b, c) show that PHP $CsPbBr_3$ has a root mean square roughness of 0.9 ± 0.4 nm before and after hBN transfer. SThM reveals an apparent lateral thermal conductivity of hBN on PHP $CsPbBr_3$ of 3.0 ± 0.3 W/(m·K), which is much higher than the bulk thermal conductivity of bare PHP $CsPbBr_3$ (0.43 ± 0.03 W/(m·K)) (Fig. 3d, e).

The influence of hBN on heat dissipation and on the reduction of thermally induced perovskite degradation was assessed under localized high-power continuous-wave laser irradiation. The laser-affected area was quantified from optical microscopy images of laser-induced surface damage (Figure 4) taken under increasing laser excitation. All images are shown with the same scale bar of 10 µm. Figures 4a and b show images of PHP $CsPbBr_3$ and hBN-covered PHP $CsPbBr_3$ films, respectively, recorded after 60 s of continuous wave irradiation with a blue (457 nm) laser at power densities of 2.4-4.8 MW/cm$^2$. For each sample, the first image serves as a reference and was obtained prior to laser exposure. With increasing laser power, a bright spot appears at the excitation position, as indicated by the arrows. The area of the bright region increases with increasing power, as quantified by the values labeled above each image. The bright spots imply that high local temperatures are generated. Although the temperature was not measured directly, a rough estimate of the maximum temperature rise at the beam center was obtained using $T_{max} = P_{abs}/(2kw\pi^{1/2})$[25], where $P_{abs}=\alpha P$ is the laser power absorbed by the perovskite ($\alpha$ = 0.45 is absorptance found using absorption coefficient ≈ $10^5$ $cm^{-1}$)[26], k is the thermal conductivity, and $w$ is the laser beamwidth equal to 252 nm. Owing to the small thickness of the perovskite layer, heat dissipation is dominated by the underlying substrate; therefore, the effective thermal conductivity of the Si-$SiO_2$ stack was considered, and a value of $k = 3$ W/(m·K) was used. This equation provides an order-of-magnitude estimate under steady-state conditions and neglects additional heat-loss mechanisms. The calculated temperatures corresponding to the applied laser powers are shown in the bottom-right corner of the optical images in Figure 4. At comparable laser intensities and corresponding temperatures, the damaged area of the PHP $CsPbBr_3$-hBN remains smaller. At 4.8 MW/$cm^2$, the damaged area in the heterostructure is 9 times smaller than the area observed in the PHP $CsPbBr_3$ sample.

To verify whether the reduced damage area in the PHP $CsPbBr_3$-hBN can be explained by improved heat dissipation, heat transfer simulations were performed. Two model systems were considered: a four-layer structure comprising a Si substrate, $SiO_2$, a perovskite layer, and air, and a five-layer structure in which an hBN layer was added on top of the perovskite layer. In both cases, the Si thickness was 1000 nm, $SiO_2$ thickness was set to 90 nm, covered with a 60 nm-thick perovskite layer and 50 nm of air to account for convective heat exchange at the top surface. For the hBN-covered structure, a 2.5 nm-thick hBN layer[27] was included on top of the perovskite. In the simulations, the PHP $CsPbBr_3$ layer, as the main light absorber, was initially set for 5 ns to 393 K, with a lateral size of 300 nm chosen to be comparable to the laser spot dimension. This temperature was chosen to correspond to the calculated local temperatures at the excitation power densities of 2.4-3.0 MW/$cm^2$, where the observable difference between bare and hBN-capped samples emerges. The specific heat capacity and thermal conductivity of the bare perovskite were obtained from SThM measurements and are comparable to previously reported values [6]. All other material parameters, including mass density, were taken from the literature (see SI Table S1). Figure 5 presents the simulated temperature diffusion and cooling dynamics of the PHP $CsPbBr_3$ films without and with an hBN capping layer. Figures 5a and b present two-dimensional cross-sectional temperature maps at different times for bare PHP $CsPbBr_3$ and PHP $CsPbBr_3$-hBN, respectively. The temperature is plotted as a function of lateral and vertical position, with dashed lines indicating the layer boundaries; the layer order is specified in Figure 5 inset. At 5 ns, a clear difference in the simulated perovskite surface temperature is observed between hBN-capped and uncapped samples. The high in-plane thermal conductivity of hBN promotes lateral heat spreading, thereby delocalizing heat accumulation and facilitating cooling. After 7 ns, the hBN-capped sample exhibits a lower temperature at the lateral edges of the heated region. This trend persists at 10 ns, where the hBN-capped structure shows a less confined hot region compared with the uncapped sample. By 50 ns, the perovskite surface temperature reaches approximately 342 K with hBN, while the uncapped film is at about 372 K.

Figure 6 summarizes the calculated maximum temperature as a function of time for both structures. In both samples, the temperature decreases monotonically after heating. Notably, the PHP $CsPbBr_3$-hBN heterostructure exhibits faster cooling than the bare PHP $CsPbBr_3$ film, highlighting the influence of hBN (see supplementary information (SI) Figure S1). The results demonstrate that the hBN capping layer provides a pathway for heat to escape, thereby enhancing heat dissipation at the perovskite layer.

**Experimental**

**Substrate preparation:** p-doped Si-$SiO_2$ (90 nm) and glass chips of 2x2 $cm^2$ and 3x3 $cm^2$, respectively, were cleaned from a protective polymer layer by ultrasonication in acetone and isopropyl alcohol, followed by 10 minutes of oxygen plasma cleaning. The Si-$SiO_2$ substrates were used for laser irradiation, heat conductivity, and SEM experiments. The glass substrates were used for ASE measurements.

**Perovskite deposition:**

The $CsPbBr_3$ perovskite solution was prepared following a reported procedure.[2] $PbBr_2$ (99.999% purity, ultradry, Alfa Aesar) and CsBr (99.999% purity, trace metal basis, Sigma-Aldrich) were dissolved in anhydrous DMSO. The molar ratio of CsBr:$PbBr_2$ was 2.75:1 for all the samples. The solution was stirred in a nitrogen-filled glovebox at 60 °C overnight. Prior to the deposition, it was filtered with a 0.2 µm filter. The perovskite layer was prepared by spin-coating the precursor solution at 4000 rpm for 120 s with an 11 s ramp. After spin-coating, the film was annealed on a hotplate at 100 °C for 20 min.

**PHP experiments:** PHP was performed with a parallel plate-based imprint system. During the heat-up of the system to the processing temperature of 150 °C for approximately 20 min, a pressure of 100 bar was applied. When the imprint temperature was reached, the pressure was kept constant for 30 min. Afterward, the system was cooled, and the pressure was released as soon as the temperature was below 25 °C.[2]

**hBN transfer:**

Commercially available CVD-grown hBN on copper (Cu) foil was provided by Grolltex Inc. The average thickness of the hBN multilayer was approximately 2.5 nm. A 1.5 µm-thick PMMA A4 (950k) support layer was spin-coated onto hBN on Cu foil, and then baked at 120 °C for 10 min. A plastic frame with a window aligned with the hBN area was attached using heat-resistant tape for improved handling. An aqueous etchant consisting of $HCl/H_2O_2$/DI water in a 1:1:10 volume ratio was used to remove the Cu substrate. The resulting hBN-PMMA frame was dried and transferred onto PHP $CsPbBr_3$ at 140 °C. The frame was subsequently removed by cutting the hBN-PMMA stack with a sharp scalpel, leaving the hBN-PMMA on the target substrate. For PMMA removal, the sample was left in chlorobenzene overnight at 60 °C under inert atmosphere.[28]

**Characterization of Materials:**

XRD with filtered Cu-Kα radiation with a wavelength of 1.5405 Å was carried out via a PANalytical instrument at a current of 40 mA and a voltage of 40 kV. SEM micrographs were taken with an SEM Zeiss SUPRA 60 at 4 kV and a working distance of 3.5 mm.

SThM combined with the 3ω-technique was employed to determine the thermal conductivity and specific heat capacity, while simultaneously acquiring surface topography and the corresponding roughness, as previously described in detail.[23,29] A resistive thermal probe (VITA-SThM probe from Bruker) served simultaneously as a local heater and a thermometer. The probe was powered by an AC excitation at two selected frequencies in the kHz range, generating periodic Joule heating at the tip and inducing a corresponding temperature oscillation of the probe. These temperature oscillations of the probe were detected using the output signals of a Wheatstone bridge, which was fed to a lock-in amplifier to selectively demodulate the third-harmonic (3ω) component at each excitation frequency. The resulting voltage components at ω and 3ω were compared to extract the local thermal response and quantitatively determine the thermal conductivity.

The SThM system used in this work is integrated into the analysis chamber of a SEM; therefore, all measurements were performed under high vacuum ($5 \times 10^{-6}$ mbar). Under these conditions, heat losses due to convection and surface degradation due to ambient gases can be excluded. During measurements, a constant low contact force was maintained between the probe and the sample to ensure stable thermal contact resistance while minimizing the risk of mechanical damage to the sample surface.

For ASE measurements the samples were excited using a pulsed laser (PowerChip NanoLaser, TEEM Photonics, France) operating at $\lambda$ = 355 nm with a pulse duration of 300 ps and a repetition rate of 1 kHz. The excitation beam was focused on the sample to a spot size of ≈ 0.25 $mm^2$. The excitation density was controlled by a neutral density filter wheel, and the laser power was monitored with a photodiode power sensor (S130VC, Thorlabs). The emission was coupled into a monochromator (Acton SP2500, Princeton Instruments, 300 lines $mm^{-1}$ grating) via a multimode optical fiber, and the spectrally dispersed signal was detected by a thermoelectrically cooled charge-coupled device camera (Princeton Instruments).

Optical images were acquired using a WiTec confocal Raman microscope (model alpha300R). Continious wave laser exposure experiments were conducted by using the same setup with a 457 nm laser at 2.4, 3.02, 3.62, and 4.82 $MW/(cm)^2$ for 60 s each. The laser spot was approximately 300 nm in diameter. The measurements were taken at room temperature (21 °C) under clean room conditions at 46 % humidity.

**Simulation**:

The heat equation was solved in Python using the NumPy package. More details are provided in the SI.

**Conclusion**

In this study, we demonstrated that the high lateral thermal conductivity of hBN improved heat dissipation at the perovskite surface. A 2.5 nm hBN was transferred onto PHP $CsPbBr_3$ thin films via a semidry process, preserving the film morphology and the ASE threshold. SThM revealed a thermal conductivity of $3.0 \pm 0.3$ W/(m·K) for an PHP $CsPbBr_3$-hBN-heterostructure, which is higher than that of the bare PHP $CsPbBr_3$ ($0.43 \pm 0.03$ W/(m·K)). Owing to the higher thermal conductivity of the capped with hBN perovskite, the laser-induced damaged area was reduced by approximately a factor of nine compared to uncapped films. This is supported by heat-diffusion simulations, which show enhanced heat spreading and faster cooling dynamics in the PHP $CsPbBr_3$-hBN-heterostructure compared to bare perovskite films. This work highlights the potential of hBN as an effective thermal management layer for next-generation perovskite-based devices.

**Author contributions**

The experiments were conceived and conducted by L.S., N.K., A.K., J.R., M.M. and M.C.L. The samples were prepared by L.S. and C.K. Simulations were performed by L.S. All authors collaborated on the interpretation of the experiments. The manuscript was written and revised by all. The work was supervised by T.R., M.M. and M.C.L.

**Conflicts of interest**

There are no conflicts to declare.

**Acknowledgments**

This project has been funded by Deutsche Forschungsgemeinschaft within TRR 404 Active-3D (project number: 528378584), the European Union's Horizon 2020 research and innovation programme under the project FOXES (951774), the German Research Foundation (DFG) through the projects Hiper-Lase (GI 1145/4-1, LE 2440/12-1 and RI 1551/18-1), as well as Thermal transport in metal-halide perovskite semiconductors under operating conditions (HE 2698/11-1), and the German Ministry of Research, Technology and Space (BMFTR) through the project NEPOMUQ (13N17112 and 13N17113). N.K. acknowledges the Alexander von Humboldt Stiftung (Humboldt Research Fellowship Programme for Postdocs). J.R. acknowledges funding through the German Ministry of Economic Affairs and Energy (BMWE) through the project KATHOGRAPH (Grant ID: 011F23219N). The authors thank Dr. Christine Hendriks (Chair of Electronic Devices at RWTH Aachen University and AMO GmbH) for careful proofreading and valuable support in preparing this manuscript. They also thank P. Grewe and Dr. U. Böttger from the Electronic Materials Research Laboratory, RWTH Aachen University, for their support with XRD measurements and data analysis.

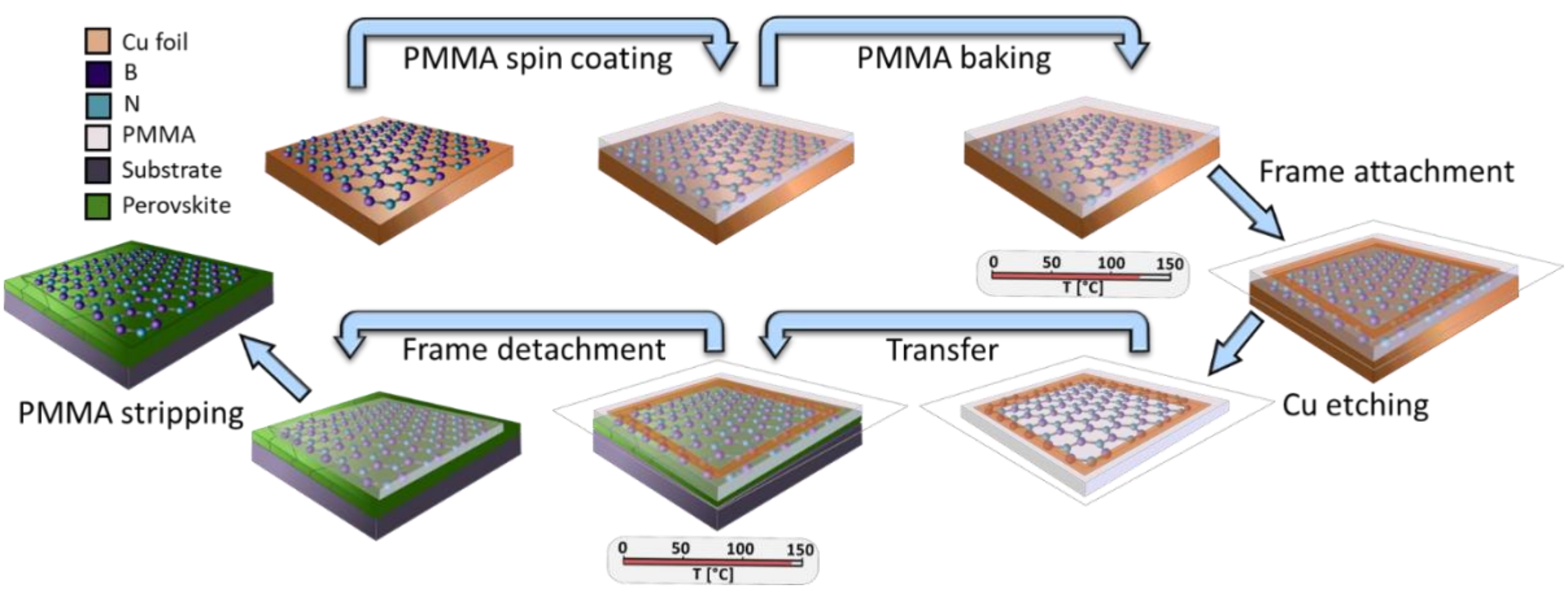

**Fig. 1** Schematic of semidry transfer of hBN onto PHP $CsPbBr_3$. Blue arrows indicate order of the process steps: Cu foil with hBN and spin-coated PMMA and its baking, attachment of the plastic frame with the heat-resistant tape, Cu etching, transfer of the stack at 140 °C, frame detachment, and PMMA removal.

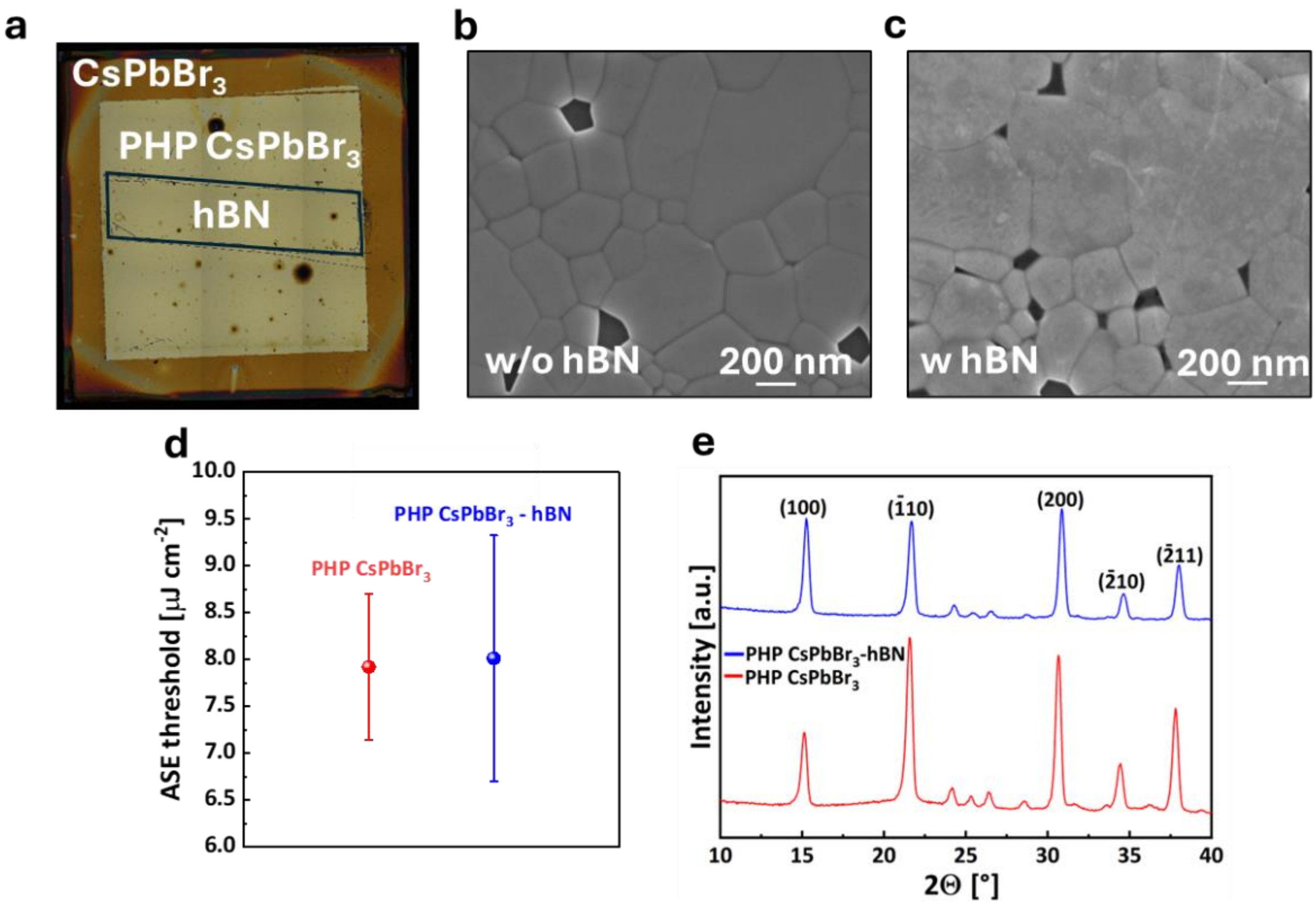

**Fig. 2** (a) Optical image of the sample after the transfer process, where the rectangular shape indicates the frame cut and defines the hBN area; (b, c) top-view SEM images of PHP $CsPbBr_3$ before and after hBN transfer, respectively; (d) ASE threshold of PHP $CsPbBr_3$ (red) and PHP $CsPBBr_3$-hBN (blue), (e) XRD spectra of PHP $CsPbBr_3$ before (red curve) and after (blue curve) hBN transfer.

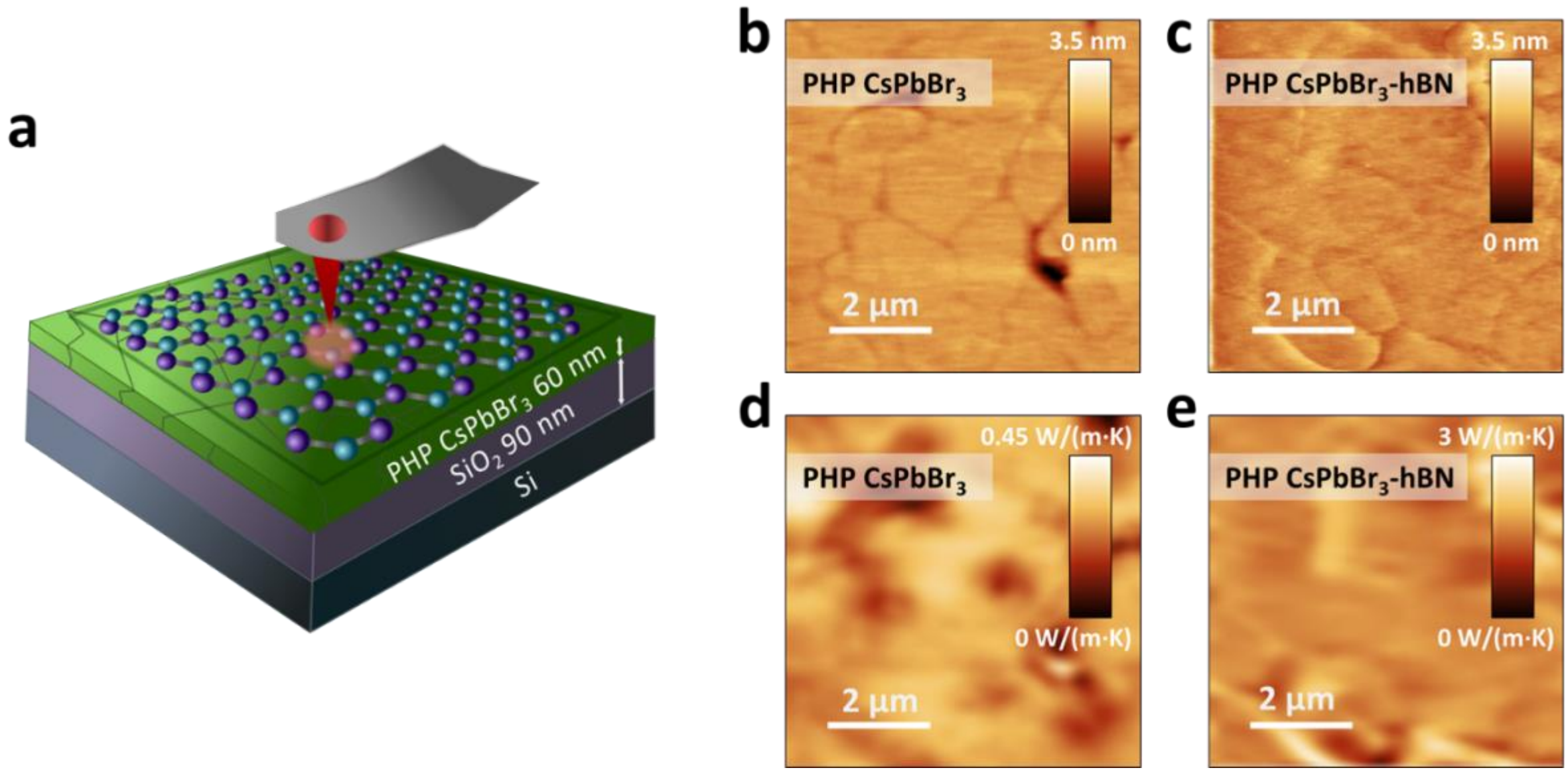

**Fig. 3** (a) Schematic of the SThM measurement, showing a temperature-sensitive probe locally probing heat dissipation in a PHP $CsPbBr_3$ thin film (60 nm) on a $SiO_2$ substrate (90 nm) with a Si layer underneath; (b, c) topography on PHP $CsPbBr_3$ and PHP $CsPbBr_3$-hBN thin film and (d, e) the corresponding thermal conductivity maps, respectively.

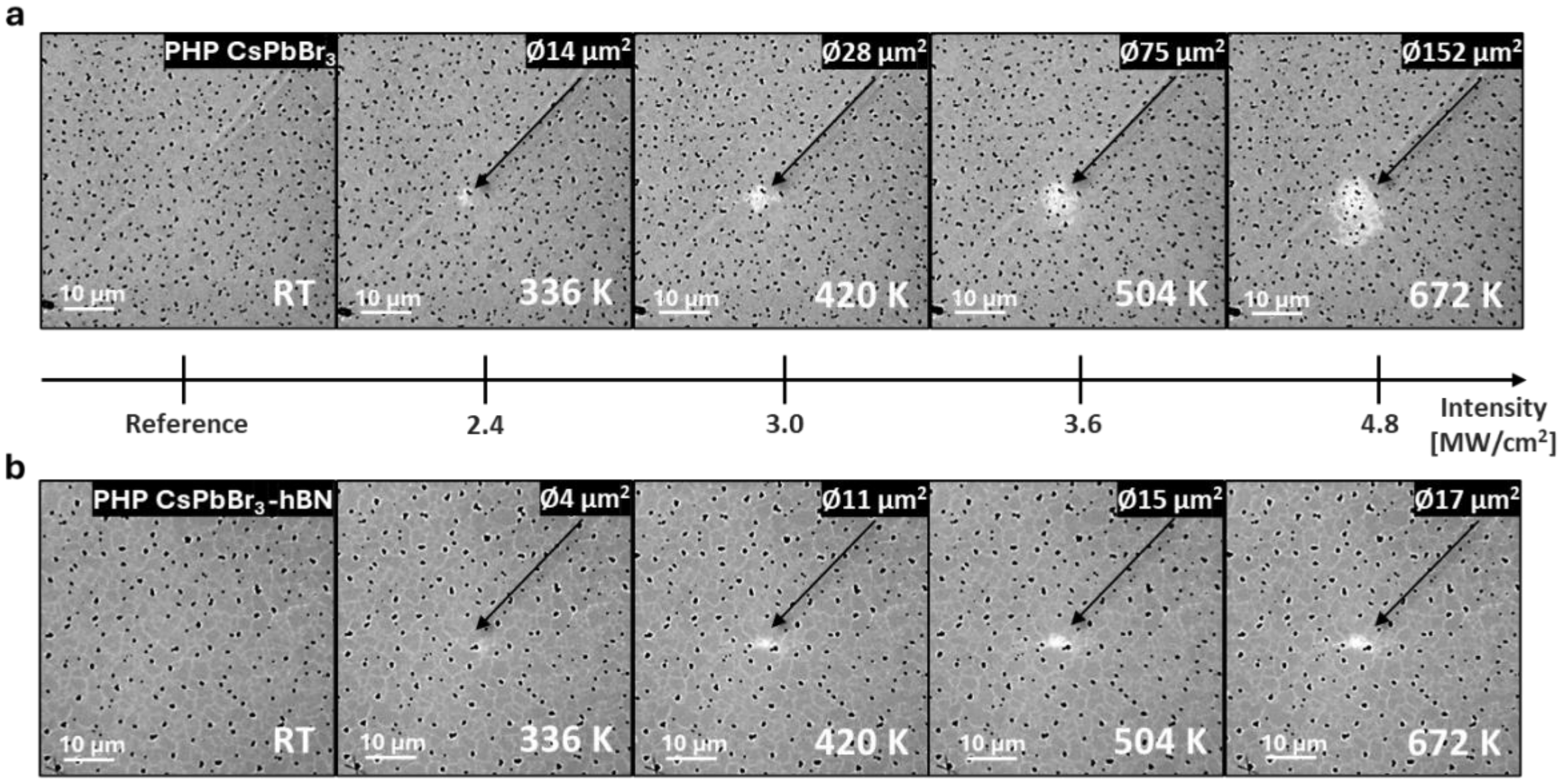

**Fig. 4.** Optical images of PHP $CsPbBr_3$ (a) and PHP $CsPbBr_3$-hBN (b) surfaces after 60 s of laser exposure at excitation powers ranging from 2.4 to 4 8 $MW/cm^2$. The black arrow indicates the laser-damaged region of the $CsPbBr_3$ surface (white spot), with the corresponding approximate area.

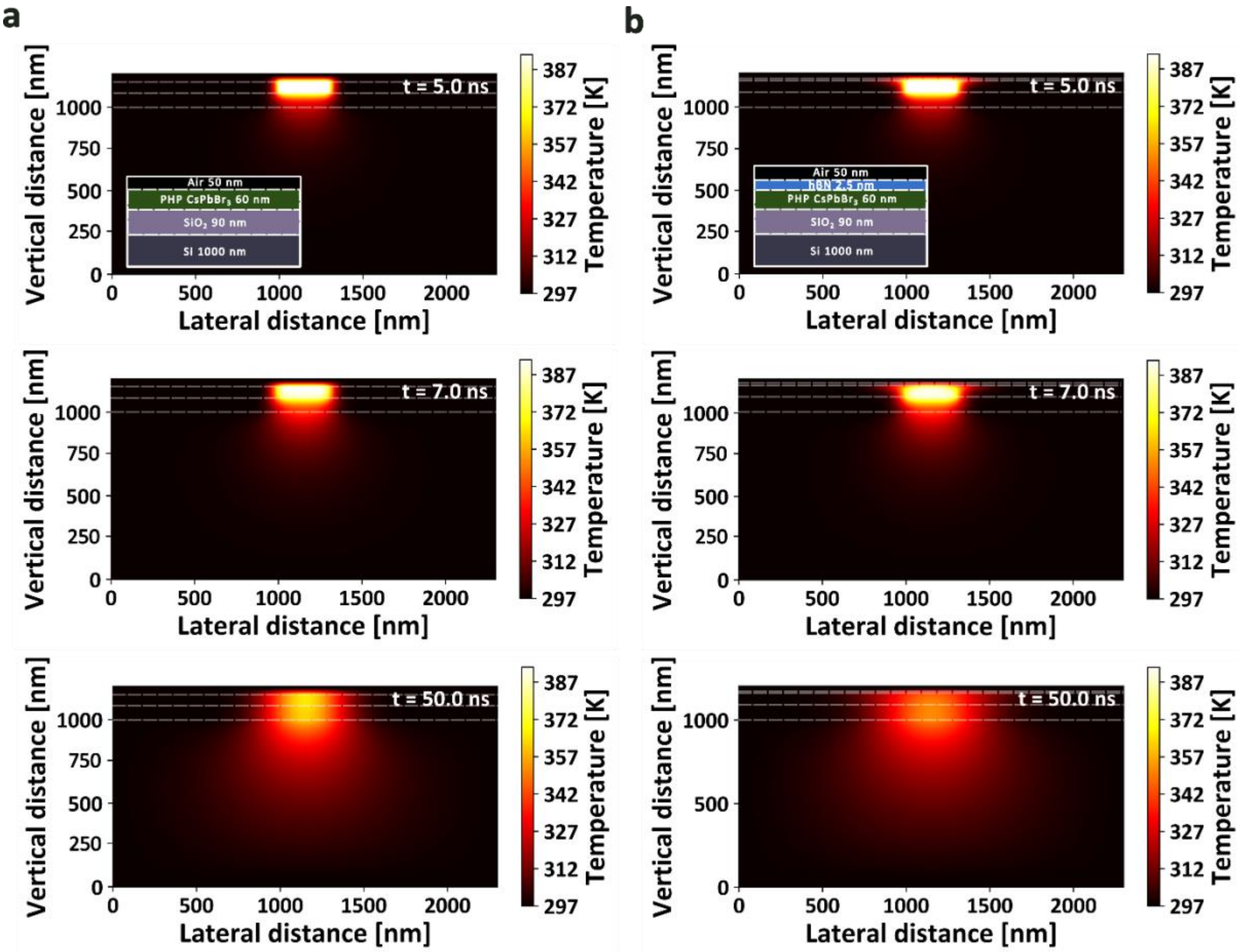

**Fig. 5** Simulation of two-dimensional heat transport as a function of time (a). Heat diffusion in the Si-$SiO_2$-PHP $CsPbBr_3$ sample. (b) Heat diffusion in the Si-$SiO_2$-PHP $CsPbBr_3$-hBN heterostructure. The white dashed lines indicate the layer thickness.

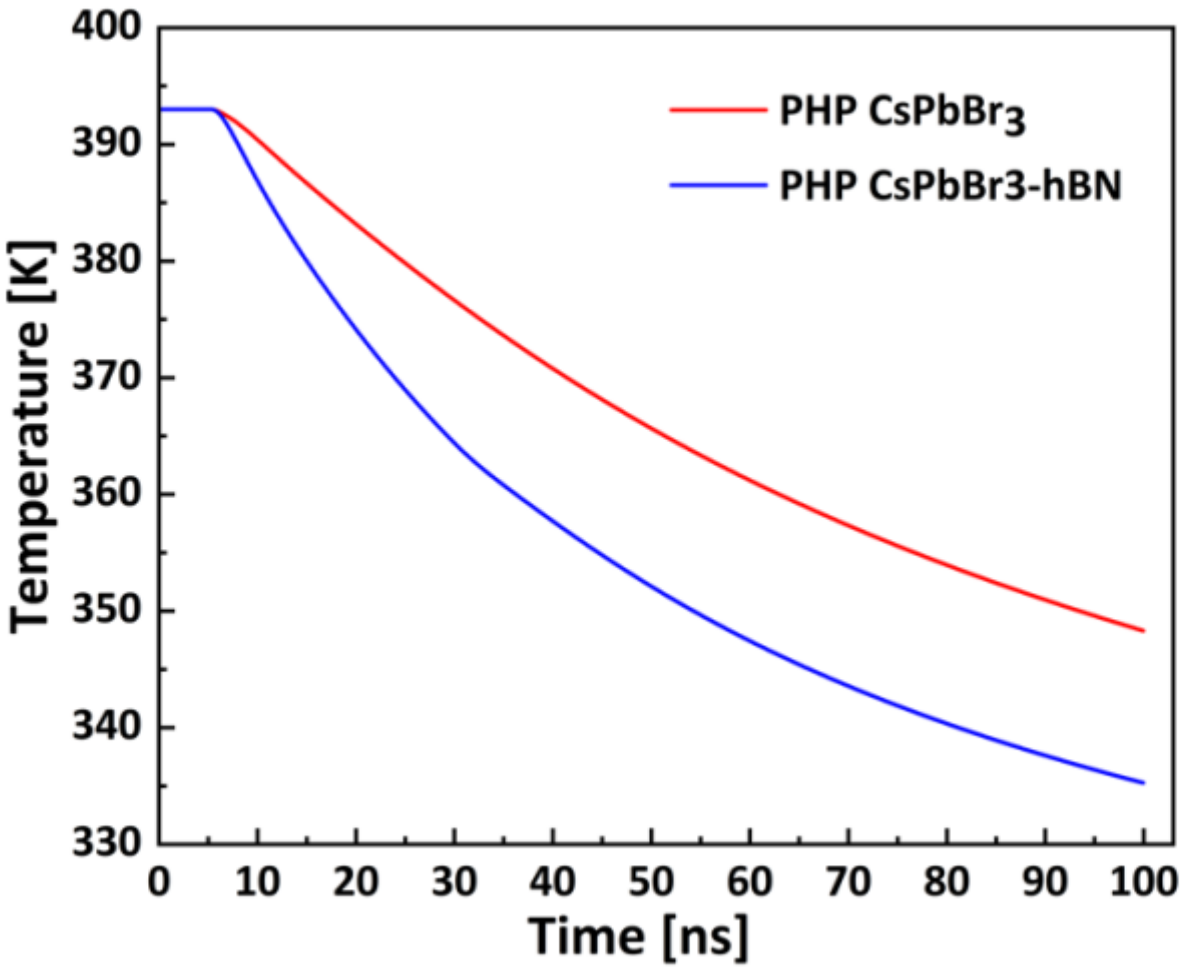

**Fig.6** Time evolution of the perovskite maximum temperature. The red and blue curves correspond to PHP $CsPbBr_3$ and PHP $CsPbBr_3$-hBN samples, respectively.

Supplementary Information

# Improved Heat Dissipation in $CsPbBr_3$-hBN Heterostructures

Liudmila Starodubtceva, [a] Ahmed Kadid, [b] Naho Kurahashi, [b] Cedric Kreusel, [b]

Jasper Ruhkopf, [ad] Sebastian Lukas, [a] Ulrich Plachetka, [d] Ralf Heiderhoff, [bc] Thomas Riedl, [bc]

Maryam Mohammadi, *[d] Max C. Lemme *[ad]

[a] Chair of Electronic Devices, RWTH Aachen University, Otto-Blumenthal-Str. 25, 52074 Aachen, Germany.

[b] Institute of Electronic Devices, University of Wuppertal, Rainer-Gruenter-Str. 21, 42119 Wuppertal, Germany.

[c] Wuppertal Center for Smart Materials & Systems (CM@S), University of Wuppertal, Rainer-Gruenter-Str. 21, 42119 Wuppertal, Germany

[d] AMO GmbH, Otto-Blumenthal-Str. 25, 52074 Aachen, Germany.

*Corresponding Authors: max.lemme@eld.rwth-aachen.de, mohammadi@amo.de

## Section 1. Simulation of thermal dissipation

Thermal simulations were performed to analyze heat generation and dissipation in perovskite-hBN heterostructures. The model geometry represents a layered stack consisting of air, a $CsPbBr_3$ perovskite thin film, an hBN overlayer (when present), and a Si-$SiO_2$ substrate. The layer thicknesses (except of 1000 nm Si) were chosen to match the experimentally studied structures.

The simulations were carried out in Python (NumPy package) using a two-dimensional transient heat-diffusion model, assuming translational invariance along the out-of-plane direction. Heat transport was considered in the plane defined by the film normal and the lateral direction, capturing both vertical heat flow into the substrate and lateral heat spreading within the layered structure.

Thermal excitation was modeled by a temperature increase of the perovskite layer. The perovskite was heated to 393 K for a duration of 5 ns, after which the system was allowed to cool through heat diffusion into air or hBN-air and Si-$SiO_2$.

Heat transport in the layered structure was described using the transient heat diffusion equation[1]:

$$\frac{\partial \mathrm{T}}{\partial \mathrm{t}} = \frac{\kappa}{\rho c_p}\left(\frac{\partial^2 T}{\partial x^2} + \frac{\partial^2 T}{\partial y^2}\right) \tag{1}$$

Where T is the temperature, ρ is the mass density, $c_p$ is the specific heat capacity, κ is the thermal conductivity tensor (see Table S1). We also applied anisotropic thermal conductivity of hBN.

Convective heat transfer to air was applied at the top surface using a Robin boundary condition, while fixed-temperature (Dirichlet at 297 K) boundary conditions were imposed at the bottom and lateral sides of the structure. A lateral width of 2300 nm was used in the simulations, with the heated central 300 nm perovskite to decrease effect of the boundaries conditions as well as make area comparable to the localized area exposed to laser beam in the real experiments.

No optical absorption or volumetric heat generation was explicitly modeled in the simulations. Thermal excitation was introduced solely through temperature increase in the perovskite layer. This approach avoids assumptions regarding absorption coefficients, optical penetration depth, or nonradiative recombination processes and allows direct comparison of heat diffusion and cooling dynamics between the layers.

The high in-plane heat conductivity of hBN is the main reason for the improved lateral heat redistribution. It is evidenced by the broadened and flattened temperature profile along the perovskite surface in the presence of hBN, indicating reduced lateral temperature gradients and enhanced in-plane heat spreading (see Fig.S1).

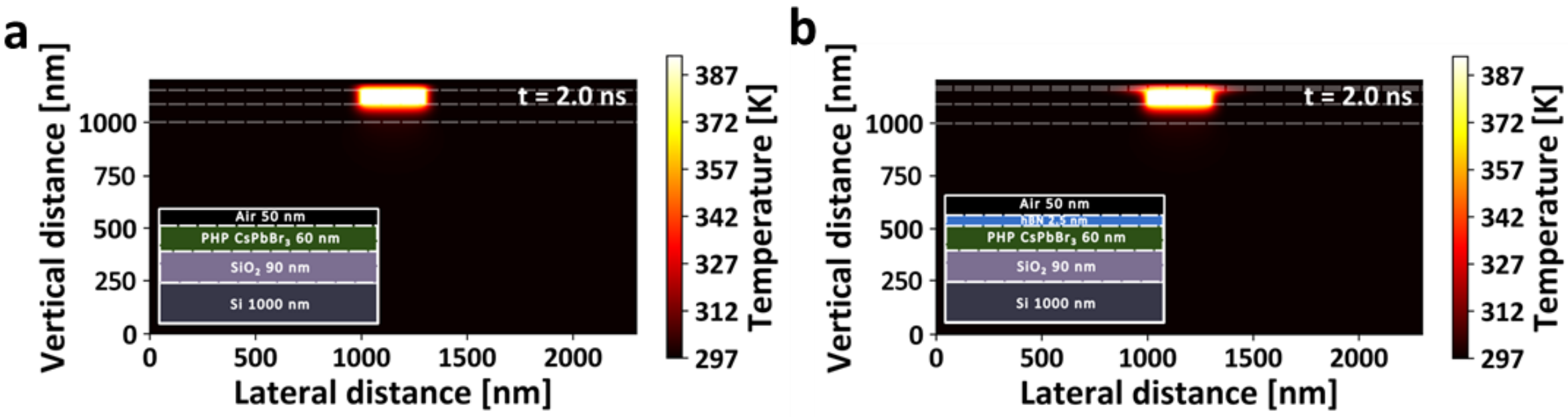

**Figure S1** Simulation of two-dimensional heat diffusion after 2 ns heating $CsPbBr_3$ at 393 K (a) from the $CsPbBr_3$ surface into air and Si-$SiO_2$. (b) Heat diffusion from the $CsPbBr_3$ surface in the presence of hBN. The white dotted lines indicate the thicknesses of the individual layers.

**Table S1**. Parameters for hBN, PHP $CsPbBr_3$, $SiO_2$, and air

| Material | hBN | PHP $CsPbBr_3$ | $SiO_2$ | Air | Si |
|---|---|---|---|---|---|
| Specific heat capacity [J/(kg·K)] | 770 | 280[2] | 680 | 920[3] | 668 |
| Mass density [$kg/m^3$] | 1900 | 4750[4] | 2170 | 1.2[3] | 2280 |
| Thermal conductivity [W/(m·K)] | 360 in-plane[5]<br>0.4 cross-plane[6] | 0.43[2] | 1.3 | 0.026[3] | 84 |

If the source is not mentioned, the following online resources were used for the material parameters:

hBN: AZO MATERIALS, https://www.azom.com/properties.aspx?ArticleID=78, (accessed 30.12.2025)

$SiO_2$: AZO MATERIALS, https://www.azom.com/properties.aspx?ArticleID=1114, (accessed 30.12.2025)

Si: AZO MATERIALS, https://www.azom.com/properties.aspx?ArticleID=1851, (accessed 12.02.2026)